\documentclass[11pt]{article}
\usepackage{amsmath,amssymb}
\usepackage[dvips]{graphicx}
\usepackage[abs]{overpic}
\usepackage{pict2e}
\newcommand{\ba}{\begin{alignat}{3}}

\newcommand{\pa}{\partial}

\newcommand{\mc}{\mathcal}

\begin{document}


\begin{flushright}
OU-HET 617
\\
November 7, 2008
\end{flushright}
\vskip3cm
\begin{center}
{\LARGE {\bf The CFT-interpolating Black Hole \\[0.2cm] in Three Dimensions}}
\vskip3cm
{\large 
{\bf Kyosuke Hotta, 
Yoshifumi Hyakutake, 
Takahiro Kubota, 
\\
Takahiro Nishinaka 
and 
Hiroaki Tanida\footnote{hotta, hyaku, kubota, nishinaka, hiroaki@het.phys.sci.osaka-u.ac.jp}}
}

\vskip1cm
{\it Department of Physics, Graduate School of Science, 
\\
Osaka University, Toyonaka, Osaka 560-0043, Japan}
\end{center}

\vskip1cm
\begin{abstract}
We present a new exact black hole solution in three dimensional Einstein gravity coupled to a single scalar field.
This is one of the extended solutions of the BTZ black hole and  has in fact $\textrm{AdS}_3$ geometries both at the spatial infinity and at the event horizon.
An explicit derivation of Virasoro algebras for $\textrm{CFT}_2$ at the two boundaries is shown to be possible \`{a} la Brown and Henneaux's calculation.
If we regard the scalar field as a  running coupling in the dual two dimensional field theory, and its flow in the bulk as the ``holographic" renormalization group flow, our black hole 
should interpolate the two $\textrm{CFT}_2$ living at the infinity and at the horizon.
Following the Hamilton-Jacobi analysis by de Boer, Verlinde and Verlinde,  
we calculate the central charges $c_{\textrm{UV}}$ and  $c_{\textrm{IR}}$ for the $\textrm{CFT}_2$ on the infinity and the horizon, respectively.
We also confirm that the inequality 
$c_{\textrm{IR}} < c_{\textrm{UV}}$ is satisfied, which is consistent with the Zamolodchikov's c-theorem.
\end{abstract}


\vfill\eject


\section{Introduction}

It has been believed that three dimensional gravity with negative cosmological constant 
is an important key to uncover some aspects of the quantum gravity.
A vacuum solution is described by three dimensional anti-de Sitter ($\textrm{AdS}_3$) geometry globally~\cite{deser},
and a black hole solution with mass and angular momentum (BTZ black hole) is constructed 
from locally $\textrm{AdS}_3$ geometry with appropriate identifications of boundaries~\cite{btz}.
All these geometries become asymptotically $\textrm{AdS}_3$, and two dimensional conformal field theory
($\textrm{CFT}_2$) is expected to exist at the boundary of the three dimensional geometry~\cite{brown}.
This is one of important examples of AdS/CFT correspondence~\cite{m},
which enables us to evaluate physical quantities in CFT from the gravity side~\cite{gkp,w}.
Thus many efforts have been directed to understand quantum nature of the three dimensional gravity
from the viewpoint of AdS/CFT correspondence.

In fact, by using the canonical formalism, Brown and Henneaux showed that general coordinate 
transformations which preserve the boundary behavior of the geometries form Virasoro algebras
for left and right movers~\cite{brown}.
Furthermore they succeeded to evaluate central extensions of these algebras and found that 
the central charges for left and right movers take the same value.
From the $\textrm{CFT}_2$ viewpoint, the globally $\textrm{AdS}_3$ corresponds to the ground state, and
the BTZ black hole does to excited states. 
Surprisingly, the macroscopic entropy of the BTZ black hole is explained by counting the number of 
degenerate states in $\textrm{CFT}_2$~\cite{as}.

The Brown-Henneaux's canonical approach has been applied to several interesting three dimensional theories.
In ref.~\cite{saida}, the gravity theory with higher derivative corrections was considered.
The values of central charges are scaled because of the higher derivative terms.
An application to the topologically massive gravity was done in ref.~\cite{hhkt},
and left-right asymmetric central charges were derived due to the gravitational Chern-Simons term.
On the other hand, the canonical approaches to the theories of Einstein gravity coupled to scalar fields were 
investigated in refs.~\cite{mz}-\cite{bt}.
Especially, it was shown in~\cite{nos} that by the canonical formalism 
the Virasoro algebras are also realized on the asymptotic boundary of the Mart\'{i}nez-Zanelli 
black hole~\cite{mz}, which has the $\textrm{AdS}_3$ geometry only at the spatial infinity.
A microscopic entropy at the spatial infinity derived by Cardy's formula was different from 
the macroscopic entropy evaluated at the horizon, and it only gives a maximum possible entropy~\cite{car,nos,park}.

The purpose of the present paper is to construct the black hole solution with a non-trivial scalar potential,
which allows the $\textrm{AdS}_3$ geometry not only at the infinity but also at the horizon.
Our black hole solution is one of extensions of the extremal BTZ black hole and no longer becomes 
$\textrm{AdS}_3$ between the spatial infinity and the horizon\footnote{A black hole solution which 
interpolates two $\textrm{AdS}_2$ is discussed in~\cite{cs}.}.
Hence, two $\textrm{CFT}_2$ should exist on the boundaries of the infinity and on the horizon with different 
central charges since effective radii of two $\textrm{AdS}_3$ are different, 
which are related to the depth of the potential.

Actually we show how to construct the Virasoro algebras for $\textrm{CFT}_2$ at two boundaries.
At the spatial infinity, we can employ the Brown-Henneaux's approach and estimate 
the values of the central charges.
Near the horizon, however, we need to impose different boundary conditions for locally $\textrm{AdS}_3$ geometries
which are preserved under the general coordinate transformations. 
Recently, for the four dimensional extremal Kerr black hole, Guica, Hartman, Song and Strominger~\cite{ghss} 
found proper boundary conditions and derived the Virasoro algebra at the horizon by taking Bardeen-Horowitz's 
near horizon limit~\cite{bh}\footnote{In refs.~\cite{car,car2}, the Virasoro algebra at the stretched horizon
was derived with the use of the canonical symplectic form. See also refs.~\cite{carlip,park2,kkp} for further discussion.}.
Since the black hole solution which we present in this paper is also extremal and the near horizon geometry 
is the same after neglecting an extra direction, it is straightforward to adapt their results to our case.
We calculate the central charges for the $\textrm{CFT}_2$ 
dual to $\textrm{AdS}_3$ at the infinity and the horizon, respectively.

Since our solution contains two fixed points which correspond to conformal field theories, 
it is interesting to investigate the renormalization group flow between them. 
According to the idea of ``holography", a change of the energy scale in the field theory is related to that of 
the radial coordinate on the gravity side~\cite{fgpw,ST,dbvv}.
At each position of the radial coordinate, two dimensional non-conformal field theory is realized on the surface.
The UV or IR region of the field theory corresponds to the spatial infinity
or horizon in the gravity theory.
In ref.~\cite{dbvv}, de Boer, Verlinde and Verlinde showed that the Hamilton-Jacobi equation for the bulk gravity 
implies the Callan-Symanzik equation for the dual field theory on the surface of the fixed radial coordinate.
The scalar fields can be identified with running couplings if the radial coordinate of the bulk can be seen as 
the cut-off scale for the dual field theory, and flows of their solutions in the bulk are understood as the 
holographic renormalization group flow~\cite{fgpw}-\cite{l}.
Generalizations to gravity theories with higher derivative 
terms are done in refs.~\cite{noo1,fms1}.

By using the Hamilton-Jacobi formalism, we derive the flow equation for our black hole solution, and calculate 
the central charges $c_{\textrm{UV}}$ and  $c_{\textrm{IR}}$ from the conformal anomaly.
It is confirmed that the inequality $c_{\textrm{IR}}<c_{\textrm{UV}}$ is satisfied
independently of the parameter of the potential, and that the c-function defined 
from the bulk gravity monotonically decreases with respect to the scale.
These results are consistent with the Zamolodchikov's c-theorem for the two dimensional field theory~\cite{z}.
The correspondence of the bulk/boundary theory is actually ensured by these computation of the central charges and the derivation of the beta function, the Callan-Symanzik equation and the c-function from the bulk gravity.
Therefore, from these observations, we conclude that our black hole solution interpolates the 
two $\textrm{CFT}_2$ at the infinity and the horizon.

Our paper is organized as follows. 
In section 2, we show the new black hole solution for the three dimensional gravity coupled to a scalar field.
In section 3, we explicitly evaluate the central charges of the Virasoro algebras at the spatial infinity and at the horizon.
In section 4, after presenting some reviews on the Hamilton-Jacobi formalism, we derive the conformal anomaly for 
the $\textrm{CFT}_2$ on the infinity and the horizon, and the Callan-Symanzik equation for the dual field theory.
The confirmation of the c-theorem is also mentioned.
Results and future discussion are summarized in section 5.
Some technical calculations are relegated in appendices A and B.

\section{The Black Hole Solution}

In this section, we discuss a black hole solution which interpolates two $\textrm{AdS}_3$ geometries
at the infinity and the horizon.
Let us start with the three dimensional Einstein gravity coupled to a scalar field:
\begin{equation}
  \mathcal{I}=\frac{1}{16\pi G_\text{N}}\int d^3x \sqrt{-G}\left[R-V(\phi)-\frac{1}{2} 
  \partial_\mu \phi \partial^\mu \phi \right],
\label{action}
\end{equation}
where $G_\text{N}$ is the Newton constant and $\mu = 0,1,2$. 
The metric and the scalar field are denoted by $G_{\mu\nu}$ and $\phi$, respectively.
The variation of the action gives the following equations of motion
\begin{alignat}{3}
  &R_{\mu\nu} - \frac{1}{2} \partial_\mu \phi \partial_\nu \phi - G_{\mu\nu} V(\phi) = 0, \notag
  \\
  &\frac{1}{\sqrt{-G}} \partial_\mu \big( \sqrt{-G} \partial^\mu \phi \big) - \frac{\partial V(\phi)}{\partial \phi} = 0,
  \label{eq:eom1}
\end{alignat}
and in order to solve these equations, we choose the BTZ-like ansatz for the metric as
\begin{alignat}{3}
ds^2&=-e^{2f(r)}dt^2+e^{2h(r)}dr^2+r^2\left(d\varphi+e^{g(r)} dt\right)^2 \label{metric}.
\end{alignat}
Substituting the ansatz into eq.~(\ref{eq:eom1}), 
we obtain five differential equations,
\begin{alignat}{3}
  0 &= V + \frac{1}{r}e^{-2h} f' + e^{-2h} f'^2 - \frac{1}{2} e^{-2f+2g-2h} r^2 g'^2 - e^{-2h}f'h' + e^{-2h} f'', \label{eq:e1}
  \\
  0 &= -3g' + rf'g' -rg'^2 + rg'h' - rg'', \label{eq:e2}
  \\
  0 &= - V + \frac{1}{r}e^{-2h} h' - e^{-2h} f'^2 + \frac{1}{2} e^{-2f+2g-2h} r^2 g'^2 + e^{-2h}f'h' - e^{-2h} f''
  - \frac{1}{2}e^{-2h}\phi'^2, \label{eq:e3}
  \\
  0 &= - V - \frac{1}{r}e^{-2h} f' - \frac{1}{2} e^{-2f+2g-2h} r^2 g'^2 + \frac{1}{r}e^{-2h} h', \label{eq:e4}
  \\
  0 &= -\frac{\partial V}{\partial \phi} + e^{-2h} \Big( \frac{1}{r}+f'-h' \Big)\phi' + e^{-2h}\phi''. \label{eq:e5}
\end{alignat}
It seems that five unknown functions $f(r)$, $g(r)$, $h(r)$, $\phi(r)$ and $V(\phi(r))$ can be determined completely
by solving the above five equations. This is not true, however.
Multiplying $\phi'$ by eq.~(\ref{eq:e5}), we obtain
\begin{alignat}{3}
  0 = - V' + e^{-2h} \Big( \frac{1}{r}+f'-h' \Big)\phi'^2 + \frac{1}{2} e^{-2h}(\phi'^2)'. \label{eq:e6}
\end{alignat}
From eqs.~(\ref{eq:e1}), (\ref{eq:e2}) and (\ref{eq:e3}), we express $g$, $V(\phi)$ and $\phi'^2$
as functionals of $f$ and $h$. 
Then we find that the equation derived by inserting these into eq.~(\ref{eq:e6}) is equivalent to eq.~(\ref{eq:e4}).

Since we have four equations among five functions, let us choose the potential energy in the form of
\begin{equation}
V(\phi)=\frac{1}{8a^4\ell^2 }\left(
-16-4\phi^2-\phi^4+32e^{-a^2+\frac{\phi^2}{4}}-16e^{-2a^2+\frac{\phi^2}{2}}+4\phi^2e^{-2a^2+\frac{\phi^2}{2}}
\right),
\label{potential}
\end{equation}
where $a$ is a dimensionless parameter\footnote{In practice, the form of the potential energy is obtained by fixing
that of the scalar field as in eq.~(\ref{eg}).}.
A shape of the potential is illustrated in fig.~1.

\begin{figure}[tb]
\begin{center}
\begin{overpic}[width=8cm,clip]{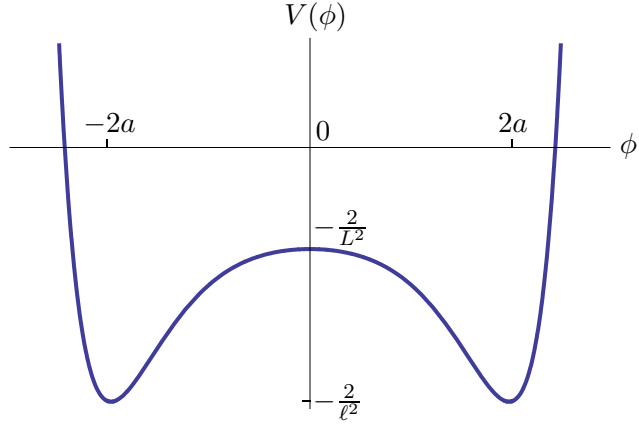}
  \put(231,97){$\phi$}
  \put(104,146){$V(\phi)$}
  \put(37,99){\line(0,1){3}}
  \put(190,99){\line(0,1){3}}
  \put(111,3){\line(1,0){3}}
  \put(28,105){$-2a$}
  \put(185,105){$2a$}
  \put(116,102){$0$}
  \put(115,66){$-\frac{2}{L^2}$}
  \put(115,0){$-\frac{2}{\ell^2}$}
\end{overpic}
\caption{Shape of the potential $V(\phi)$.}
\end{center}
\end{figure}

Extrema of the potential are realized at $\phi=0$ and $\phi=\pm 2a$, 
and the values of the potential energy become negative, $V(0)=-2/L^2$ and $V(\pm 2a)=-2/\ell^2$.
Here we defined
\begin{alignat}{3}
  L \equiv \frac{a^2 \ell}{1-e^{-a^2}}, \label{eq:L}
\end{alignat}
which satisfies $\ell < L$ when $0 < a$.
Note that there are constant scalar solutions, $\phi(r)=0$ and $\phi(r)=2a$.
In these cases, the geometries reduce to extremal BTZ black holes which asymptotically become AdS$_3$ with
the radius $L$ and $\ell$.
Therefore it is expected that a solution which interpolate between $\phi=0$ and $\phi=2a$ will 
generate the $\textrm{CFT}_2$-interpolating black hole.
In fact, it is possible to solve four equations and obtain $f(r)$, $g(r)$, $h(r)$ and $\phi(r)$ which
interpolate between $\phi(\infty)=0$ and $\phi(r_0)=2a$,
\begin{align}
e^{2f(r)}&=\frac{r^2}{a^4\ell^2}\left(e^{-a^2r_0^2/r^2}-e^{-a^2}\right)^2, \notag \\
e^{2h(r)}&=\frac{a^4\ell^2}{r^2}\left[1-e^{a^2\left(r_0^2/r^2-1\right)}\right]^{-2}, \notag \\
e^{g(r)}&=\frac{1}{a^2\ell}\left(1-e^{-a^2r_0^2/r^2}\right), \label{eg} \\
\phi(r)&=2a\frac{r_0}{r}. \notag
\end{align}
Here we focus our attention on the region $0\leq\phi\leq 2a$, that is, $r_0 \leq r$, and
$\varphi$ is the angular coordinate with the periodicity $2\pi$.
Notice that when $a=0$, the solution becomes the extremal BTZ black hole.

The thermodynamic properties of the black hole are evaluated at the horizon $r=r_0$. 
The temperature is obtained by the inverse of the periodicity of the Euclidean time,
\begin{alignat}{3}
  T &= \frac{1}{2\pi} e^{-h} \frac{d e^f}{dr} \Big|_{r=r_0} = 0.
\end{alignat}
Therefore the solution corresponds to an extremal black hole.
The Bekenstein-Hawking entropy is estimated by the area of the horizon as
\begin{alignat}{3}
  S_\text{BH} = \frac{A_\text{H}}{4 G_\text{N}} = \frac{\pi r_0}{2G_\text{N}}. \label{eq:BH}
\end{alignat}

In the following, we see that this solution represents the extremal black hole
which enables us to have two $\textrm{AdS}_3$ geometries at the spatial infinity and the horizon.
First let us investigate behaviors of the geometry around $r=\infty$ by taking the limit of $r_0 \ll r$.
Then the solution (\ref{eg}) approaches
\begin{alignat}{3}
  ds^2 &\sim - \frac{r^2}{L^2} \Big(1 - \frac{2 L r_0^2}{\ell r^2}  \Big) dt^2 
  + \frac{L^2}{r^2} \Big( 1 + \frac{2 e^{-a^2}L r_0^2}{\ell r^2} \Big) dr^2 +
  r^2 \Big(d\varphi + \frac{r_0^2}{\ell r^2} dt \Big)^2 \label{asymptotic} 
  \\
  &\sim - \frac{r^2}{L^2}dt^2 + \frac{L^2}{r^2}dr^2 +r^2d\varphi^2, \notag
\end{alignat}
where $L$ is defined in eq.~(\ref{eq:L}).
This is just the $\textrm{AdS}_3$ geometry whose radius is $L$.

On the other hand, in order to find the near-horizon geometry of (\ref{metric}), we have to put 
\begin{equation}
t=e^{a^2}\frac{t'}{\epsilon}, \,\,\,\,\,\,\,\,
r=r_0+\epsilon r',
\,\,\,\,\,\,\,\,
\varphi=\varphi'-\frac{e^{a^2}-1}{a^2\ell}\frac{t'}{\epsilon},
\label{nearh}
\end{equation}
and then in the limit $\epsilon\rightarrow0$ we obtain
\begin{equation}
  ds^2\sim -\frac{4}{\ell^2}r'^2 dt'^2 +\frac{\ell^2}{4}\frac{dr'^2}{r'^2} +r_0^2\left(
  d\varphi'-\frac{2}{r_0\ell}r' dt' \right)^2. \label{hmetric}
\end{equation}
In fact, we can confirm that this is the same as the near horizon limit of the extremal BTZ black hole whose radius is $\ell$.
This near horizon behavior follows from the extremality of the black hole~\cite{bh}.

When the radial coordinate $r$ is not around the spatial infinity $r=\infty$ or around the horizon $r=r_0$, 
the space-time is not $\textrm{AdS}_3$.
Therefore our solution interpolates two $\textrm{AdS}_3$ geometries at the spatial infinity and the horizon.

\section{Virasoro Algebras at the Spatial Infinity and the Horizon}

\subsection{Review of Brown-Henneaux's canonical formalism}

The solution constructed in the previous section approaches $\textrm{AdS}_3$ geometry at $r=\infty$ and $r=r_0$,
and it is expected that $\textrm{CFT}_2$ is realized at each boundary.
In this section we explicitly show how to construct the Virasoro algebras at these critical points
by using the canonical formalism.

In order to construct the Virasoro algebras at the boundary of $\textrm{AdS}_3$ geometry,
let us briefly review the Brown-Henneaux's canonical formalism.
The notations employed in this section are the same as in ref.~\cite{hhkt}.
The ADM decomposition of the three dimensional metric is expressed as
\begin{equation}
ds^2=-N^2dt^2+g_{ij}\left(dx^i+N^idt\right)\left(dx^j+N^jdt\right). \label{eq:ADM}
\end{equation}
By applying the ADM decomposition of the metric, the Lagrangian becomes
\ba
  \mc{L} &= \sqrt{g} N \big( R^{(2)} - V(\phi) + K^{ij}K_{ij} - K^2 \big)
  + \frac{\sqrt{g} }{2N} (\dot \phi - N^i \pa_i \phi)^2 
  - \frac{\sqrt{g} N}{2} \pa_i \phi \pa^i \phi, \label{eq:lag}
\end{alignat}
where the dot represents the derivative with respect to the time coordinate $t$, and 
$K_{ij} = \frac{1}{2N}(\dot g_{ij} - 2 \nabla_{(i} N_{j)})$.
$R^{(2)}$ stands for the scalar curvature made from the two dimensional metric $g_{ij}$.
Momenta conjugate to $g_{ij}$ and $\phi$ are given by $\pi^{ij} = \sqrt{- g} (K^{ij} - g^{ij} K)$ and
$\pi_\phi = \frac{1}{N} \sqrt{- g} (\dot \phi - N^i \pa_i \phi)$.
Up to total derivative terms, the Hamiltonian density is expressed as
\begin{align}
  \mc{H}_\text{D}[\xi] &= \pi^{ij} \dot g_{ij} + \pi_\phi \dot \phi - \mc{L} \notag
  \\
  &= \xi^0 \Big\{ \tfrac{1}{\sqrt{g}} \big( (\pi_{ij})^2-(\pi^i_i)^2+\tfrac{1}{2}\pi_\phi^2 \big)
  +\sqrt{g} \big( V(\phi)-R^{(2)}+\tfrac{1}{2}(\partial_i\phi)^2 \big) \Big\} \label{eq:ham}
  \\
  &\quad\,
  + \xi^i \Big\{ -2 \sqrt{g} \nabla_j \Big(\tfrac{1}{\sqrt{g}}\pi_i^j \Big) + \pi_\phi\partial_i\phi \Big\}. \notag
\end{align}
Here we introduced $\xi^0$ and $\xi^i$ which are related to ``the Killing vector" $\bar{\xi}$ via 
\begin{alignat}{3}
(\xi ^{0}, \xi ^{r}, \xi^{\varphi})
 =(N{\bar \xi}^{t}, {\bar \xi}^{r}+N^{r}{\bar \xi}^{t}, 
 {\bar \xi}^{\varphi}+N^{\varphi}{\bar \xi}^{t}),
 \label{eq:localcoordinate}
\end{alignat}
and we choose $\bar{\xi}=(1,0,0)$ for the Hamiltonian density.
The Hamiltonian is given by integrating over two dimensional 
spaces with an surface term $Q[\xi]$\cite{RT},
\begin{equation}
H[\xi]=\int d^2x \, \mathcal{H}_\text{D}[\xi] +Q[\xi].
\end{equation}
The term $Q[\xi]$ should be added so as to cancel surface variations of the Hamiltonian (\ref{eq:ham}),
and an explicit form of the variation is given by
\begin{align}
  \delta Q[\xi] &= \delta Q^G[\xi] + \delta Q^\phi[\xi] , \notag
  \\
  \delta Q^G[\xi]&=
  \int d\varphi\left[\sqrt{g} S^{ijkr}( \xi ^{0}\nabla_{k}\delta  g_{ij} - 
  \nabla_{k} \xi ^{0} \delta  g_{ij})+ (2\xi ^{i}\pi ^{jr} -\xi ^{r}\pi ^{ij}) \delta g_{ij}
  +2\xi _{i}\delta {\pi ^{ir}} 
  \right ], \label{Qg}
\\
  \delta Q^\phi[\xi]&=-\int d\varphi\left(
  \xi^r\pi_\phi\,\delta\phi+\sqrt{g}\xi^0\partial^r\phi\,\delta\phi \right). \label{Qphi}
\end{align}
Here $S^{ijkl}$ is defined by $S^{ijkl}=\frac{1}{2}\left (g^{ik}g^{jl}+g^{il}g^{jk}-2g^{ij}g^{kl} \right )$.
As is clear from the expression, $\delta Q^G[\xi]$ is the contribution from the metric and 
$\delta Q^\phi[\xi]$ is from the scalar field.

The algebraic structure of symmetric transformation group is obtained by the
Poisson bracket algebra of the Hamiltonian generator $H[\xi]$:
\begin{align}
 \left\{H[\xi ],H[\eta
 ]\right\}_{\text{P}}=H\big[[\xi,\eta]\big]+K[\xi,\eta] , 
 \label{p-b}
\end{align}
where $K[\xi,\eta]$ is a possible central extension. 
The Dirac bracket $\left \{ Q[\xi ], Q[\eta ] \right \}_{\text{D}}$
gives a surface deformation of $Q[\xi]$ with respect to
$Q[\eta]$, i.e., $\delta_\eta Q[\xi]=\left\{Q[\xi ],Q[\eta ]\right\}_\text{D}$.
The charge $Q[\xi]$ forms a conformal group together with the central extension 
$\left \{ Q[\xi ], Q[\eta] \right \}_{\text{D}}
=Q \big[ [ \xi,\eta ] \big ] + K[\xi,\eta] $, and we immediately get
$\delta _{\eta} Q[\xi] =Q\big[[\xi,\eta]\big]+K[\xi,\eta]$.
If we set $Q\big[[\xi,\eta]\big]=0$ for a vacuum ($\phi=0$ or $\phi=2a$),  the evaluation 
of the central charge reduces to 
\begin{align}
 K\big[ \xi,\eta \big]=\delta_\eta Q[\xi]. 
 \label{central-ext}
\end{align}
Namely, we only have to substitute  ``the Killing vector" which preserves each geometry into (\ref{Qg}) 
and (\ref{Qphi}) in order to calculate the central charges.

\subsection{Central charges at the spatial infinity}

Near the infinity ($r\rightarrow\infty$) the space-time (\ref{asymptotic}) can allow the following behavior of the solution,
\begin{align}
 &G_{tt}=-\frac{r^2}{L ^2}+\mathcal{O}(1),
 \,\,\,\,\,\,\,\,\,\,\,\,
 G_{tr}=\mathcal{O}(r^{-3}),
 \,\,\,\,\,\,\,\,\,\,\,\,
 G_{t\varphi}=\pm\mathcal{O}(1), 
 \notag \\
 &G_{rr}=\frac{L ^2}{r^2}+\mathcal{O}(r^{-4}),
 \,\,\,\,\,\,\,\,\,\,\,\,
 G_{r\varphi}=\mathcal{O}(r^{-3}),
 \,\,\,\,\,\,\,\,\,\,\,\,
 G_{\varphi\varphi}=r^2+\mathcal{O}(1),\notag\\
&\phi=\frac{2ar_0}{r}+\mathcal{O}(r^{-1}),
\label{hypersurface bd}
\end{align}
where $L$ is the radius of the AdS$_3$ defined in eq.~(\ref{eq:L}).
When we define $x^\pm =\frac{t}{L} \pm \varphi$, ``the Killing vector" is calculated from (\ref{hypersurface bd}),
\begin{align}
  \bar{\xi}^{\pm t}_n = \frac{L}{2} e^{inx^\pm} \left(1 - \frac{L^2 n^2}{2r^2} \right), \quad
  \bar{\xi}^{\pm r}_n = -i \frac{nr}{2} e^{inx^\pm}, \quad
  \bar{\xi}^{\pm \varphi}_n = \pm \frac{1}{2} e^{inx^\pm} \left(1 + \frac{L^2 n^2}{2r^2} \right),
  \label{eq:Killing}
\end{align}
and $\xi^\pm_m = \bar{\xi}^{\pm \mu}_m \partial_\mu$.
We often call ``$+$" left and ``$-$" right.
It is found from (\ref{hypersurface bd}) that the canonical variables behave as
\begin{align}
&g_{rr}=\frac{L ^2}{r^2}+\mathcal{O}(r^{-4}),
\,\,\,\,\,\,\,\,\,\,\,\,
g_{r\varphi}=\mathcal{O}(r^{-3}),
\,\,\,\,\,\,\,\,\,\,\,\,
g_{\varphi\varphi}=r^2 +\mathcal{O}(1),
\notag
\\
 &N=\frac{r}{L}+\mathcal{O}(r^{-1}),
 \,\,\,\,\,\,\,\,\,\,\,\,
 N^{r}=\mathcal{O}(r^{-1}),
 \,\,\,\,\,\,\,\,\,\,\,\,
 N^\varphi=\pm\mathcal{O}(r^{-2}),
\notag
\\
&\pi^{rr}=\mathcal{O}(r^{-1}),
 \,\,\,\,\,\,\,\,\,\,\,\,
 \pi^{r\varphi}=\mathcal{O}(r^{-2}),
 \,\,\,\,\,\,\,\,\,\,\,\,
 \pi^{\varphi\varphi}=\mathcal{O}(r^{-5}),
\notag\\
&\phi=\frac{2ar_0}{r}+\mathcal{O}(r^{-1}),
\,\,\,\,\,\,\,\,\,\,\,\,
\pi_\phi=\mathcal{O}(r^{-4}). \label{eq:infbc}
\end{align}
The explicit calculations are shown in the appendix \ref{appb1}.

The computation of the term $\delta_\eta Q^G[\xi]$ becomes
\begin{alignat}{3}
  \frac{1}{16\pi G_\text{N}} \delta_{\eta=\xi^+_n} Q^G[\xi=\xi^+_m]
  &= \frac{1}{16\pi G_\text{N}} \oint_{r=\infty} d\varphi\left[\sqrt{g} S^{ijkr}( \xi ^{0}\nabla_{k}\delta_\eta  g_{ij} - 
  \nabla_{k} \xi ^{0} \delta_\eta  g_{ij}) + 2\xi _{i}\delta_\eta {\pi ^{ir}} \right ] \notag
  \\
  &= -i \frac{L}{8 G_\text{N}}m^3\,\delta_{m+n,0} -i \frac{ (3 + e^{-a^2}) r_0^2}{8 G_\text{N} \ell}m\,\delta_{m+n,0},
\end{alignat}
and the variation $\delta_\eta Q^\phi[\xi]$ does
\begin{alignat}{3}
  \frac{1}{16\pi G_\text{N}}\delta_{\eta=\xi^+_n}Q^\phi[\xi=\xi^+_m]
  &=-\frac{1}{16\pi G_\text{N}} \oint_{r=\infty} d\varphi\sqrt{g} \xi^0\partial^r\phi\,\delta_\eta\phi \notag
  \\
  &=-i\frac{a^2r_0^2}{8G_\text{N} L}m\,\delta_{m+n,0}.
\end{alignat}
Combining these results and using the definition of $L$, we find that the central extension of the Virasoro algebra is given by
\begin{alignat}{3}
  \frac{1}{16\pi G_\text{N}} \delta_{\eta=\xi^+_n} Q[\xi=\xi^+_m]
  &= -i \frac{L}{8 G_\text{N}}m^3\,\delta_{m+n,0} -i \frac{ r_0^2}{2 G_\text{N} \ell}m\,\delta_{m+n,0}.
\end{alignat}
A similar calculation shows that
\begin{alignat}{3}
  \frac{1}{16\pi G_\text{N}} \delta_{\eta=\xi^-_n} Q[\xi=\xi^-_m]
  &= -i \frac{L}{8 G_\text{N}}m^3\,\delta_{m+n,0}.
\end{alignat}
From the cubic term in $m$ we conclude that left and right Virasoro algebras surely live 
at the infinity $(r\rightarrow\infty)$ and two central charges take the same value,
\begin{equation}
c_{\textrm{UV}}
=\frac{3L}{2G_\text{N}}.
\label{cuv}
\end{equation}
The linear dependences on $m$ indicate the excitations of left and right zero modes of the Virasoro algebras, 
$2L_0$ and $2\bar{L}_0$.
Thus the geometry which we are considering corresponds to
\begin{alignat}{3}
  L_0^\text{UV} = \frac{r_0^2}{4G_\text{N} \ell}, \qquad \bar{L}_0^\text{UV} = 0, \label{eq:mlinear}
\end{alignat}
and only the left moving modes are excited. In this sense, we call this geometry chiral.
The sign of $G_{t\varphi}$ determines which modes are excited or not.
In our solution (\ref{asymptotic}) the sign of $G_{t\varphi}$ is plus, and left modes are excited.
The field theory dual to extremal black holes specified the sign of the angular momentum 
is chiral even at the infinity~\cite{ghss}.

Mass and angular momentum are estimated by choosing the Killing vectors as
a time translation $\bar{\xi}=(1,0,0)$ and a rotation $\bar{\xi}=(0,0,1)$.
Because they are related to the horizon radius $r_0$, we take the shift $r_0\rightarrow r_0+\delta r_0$ 
as an explicit deformation of $\delta$ in order to know these quantities.
With the use of  $\delta g_{rr}=\frac{2a^6e^{-a^2}\ell^2\delta (r_0^2)}{(1-e^{-a^2})^3r^4}$ and 
$\delta \phi=\frac{2a\delta r_0}{r}$, (\ref{Qg}) and  (\ref{Qphi}) become
\begin{equation}
  \delta M = \frac{1}{16\pi G_\text{N}}
  \oint_{r=\infty} d\varphi\sqrt{g}
  \left( S^{ijkr}\xi ^{0}\nabla_{k}\delta  g_{ij} 
  -\xi^0\partial^r\phi\,\delta\phi\right)
  =\frac{\delta (r_0^2)}{4G_N\ell L},
\end{equation}
for $\bar{\xi}=(1,0,0)$.
In a similar way, we calculate
\begin{equation}
  \delta J = \frac{1}{16\pi G_\text{N}}
  \oint_{r=\infty} d\varphi\,(2\xi_i\delta\pi^{ir})
  =\frac{\delta (r_0^2)}{4G_\text{N}\ell},
\end{equation}
for $\bar{\xi}=(0,0,1)$.
Therefore the mass and the angular momentum become
\begin{equation}
  M_\text{UV} L = J_\text{UV} = \frac{r_0^2}{4G_\text{N}\ell},
\end{equation}
which satisfies the extremal relation and consistent with eq.~(\ref{eq:mlinear}).
Then using the Cardy's formula for the chiral $\textrm{CFT}_2$ at the infinity, we obtain the entropy
\begin{equation}
  S_\text{UV}=2\pi\sqrt{\frac{c_{\textrm{UV}}L_0^\text{UV}}{6}}
  =\frac{\pi r_0}{2G_\text{N}}\sqrt{\frac{L}{\ell}}. \label{eq:CardyUV}
\end{equation}

\subsection{Central charges at the event horizon}

The near horizon limit was given in eq.~(\ref{nearh}).
Since this does not cover the full space-time, 
we first transform  the metric (\ref{hmetric}) to
\begin{equation}
ds^2=\frac{\ell^2}{4}\left[
-(1+\tilde{r}^2)d\tilde{t}^2+\frac{d\tilde{r}^2}{1+\tilde{r}^2}
+\left(\frac{2r_0}{\ell}d\tilde{\varphi}-\tilde{r}d\tilde{t}\right)^2
\right],
\label{tilde}
\end{equation} 
where we have defined~\cite{bh}
\begin{align}
 t'&=\frac{\ell^2}{4}\frac{\sqrt{1+\tilde{r}^2}\sin\tilde{t}}{r'},\notag\\
r'&=\frac{\ell}{2}\left(\sqrt{1+\tilde{r}^2}\cos\tilde{t}+\tilde{r}\right),\notag\\
\varphi'&=\tilde{\varphi}+\frac{\ell}{2r_0}\log\left|\frac{\cos\tilde{t}+\tilde{r}\sin\tilde{t}}{1+\sqrt{1+\tilde{r}^2}\sin\tilde{t}}\right|.
\end{align}
Notice that $0\leq\tilde{\varphi}\leq 2\pi$ at $\tilde{t}=t'=0$ due to $0\leq\varphi'\leq 2\pi$.
This geometry has an isometry group of 
$SL(2,\mathbb{R})_R \times SL(2,\mathbb{R})_L$.

We are interested in the near horizon region where $\tilde{r}\rightarrow\infty$. 
According to~\cite{ghss} we assume the boundary condition
\begin{align}
 &G_{\tilde{t} \tilde{t}}=-\frac{\ell^2}{4}+\mathcal{O}(\tilde{r}^{-1}),
 \,\,\,\,\,\,\,\,\,\,\,\,
 G_{\tilde{t}\tilde{r}}=\mathcal{O}(\tilde{r}^{-2}),
 \,\,\,\,\,\,\,\,\,\,\,\,
 G_{\tilde{t}\tilde{\varphi}}=-\frac{\ell r_0\tilde{r}}{2}+\mathcal{O}(1), 
 \notag \\
 &G_{\tilde{r}\tilde{r}}=\frac{\ell ^2}{4(1+\tilde{r}^2)}+\mathcal{O}(\tilde{r}^{-3}),
 \,\,\,\,\,\,\,\,\,\,\,\,
 G_{\tilde{r}\tilde{\varphi}}=\mathcal{O}(\tilde{r}^{-1}),
 \,\,\,\,\,\,\,\,\,\,\,\,
 G_{\tilde{\varphi}\tilde{\varphi}}=r_0^2+\mathcal{O}(1),\notag\\
&\phi=2a.
\label{bchorizon}
\end{align}
Here we have supposed a stronger  condition of $G_{\tilde{t} \tilde{t}}$ than that in~\cite{ghss} because our near horizon geometry (\ref{tilde}) is exact $\textrm{AdS}_3$, not warped  $\textrm{AdS}_3$.
Then ``the Killing vector" to the first order
\begin{equation}
\bar{\xi}^{\tilde{t}}_n=0,
\,\,\,\,\,\,\,\,\,\,\,\,
\bar{\xi}^{\tilde{r}}_n=-in\tilde{r}e^{in\tilde{\varphi}},
\,\,\,\,\,\,\,\,\,\,\,\,
\bar{\xi}^{\tilde{\varphi}}_n=e^{in\tilde{\varphi}}, \label{eq:Killing2}
\end{equation} 
is obtained.
In addition to this, there is an $U(1)$ isometry which is given by $\bar{\xi}=(1,0,0)$.

Since $\phi$ is constant, the scalar contribution (\ref{Qphi}) vanishes. 
For the contribution from the gravity, we need to employ not 
eq.~(\ref{Qg}) but the covariant formalism in refs.~\cite{BB,BC},
since the fluctuations $\delta G_{\mu\nu}$ of the metric is not sub-leading. 
In the case of the Kerr geometry, this kind of prescription was applied in ref.~\cite{ghss}.
After some calculations, which is explained in the appendix \ref{appb2}, the central extension of the Virasoro algebra
is given by
\begin{alignat}{3}
  \frac{1}{16\pi G_\text{N}}\delta_{\eta=\xi_n} Q^C[\xi=\xi_{m}]
  &=- \frac{i}{12}\frac{3\ell}{2G_\text{N}}m^3\delta_{m+n,0} 
  - \frac{ir_0^2}{2G_\text{N}\ell}m\delta_{m+n,0}\,\,. \label{centralhorizon}
\end{alignat}
This means the existence of the Virasoro algebra with the central charge 
\begin{equation}
c_{\textrm{IR}}
=\frac{3\ell}{2G_\text{N}}.
\label{cir}
\end{equation}

Finally we present the expression of the mass and angular momentum from the value of (\ref{Qg}) and (\ref{Qphi}) at $\tilde{r}\rightarrow\infty$.
If we shift it as $r_0\rightarrow r_0+\delta r_0$, the canonical variables except for $g_{\tilde{\varphi}\tilde{\varphi}}=r_0^2+2r_0\delta r_0$ and $N^{\tilde{\varphi}}=-\frac{\ell\tilde{r}}{2r_0}+\frac{\ell\tilde{r}}{2r_0^2}\delta r_0$ are unchanged and the quantities $\delta M$ and $\delta J$ are calculable as deformations with respect to the shift of the horizon radius.
The angular momentum is obtained by setting $\bar{\xi}=(0,0,1)$ 
in (\ref{Qg}) and (\ref{Qphi}),
\begin{equation}
\delta J = \frac{1}{16\pi G_\text{N}}\oint_{\tilde{r}=\infty} d\tilde{\varphi}
\left(2\xi ^{i}\pi ^{j\tilde{r}} \delta g_{ij}\right)
=\frac{\delta(r_0^2)}{4G_\text{N}\ell}.
\end{equation}
Furthermore the charge with respect to $\bar{\xi}=(1,0,0)$ becomes zero. This means that 
$\bar{L}^\text{IR}_0 = M_\text{IR}\ell - J_\text{IR}=0$.
From this, the mass and the angular momentum defined at the horizon turn out to be
\begin{equation}
M_\text{IR}\ell=J_\text{IR}=\frac{r_0^2}{4G_\text{N}\ell}.
\end{equation}
From these the Cardy's entropy for the chiral $\textrm{CFT}_2$ at the horizon is given by
\begin{equation}
  S_\text{IR} = 2\pi\sqrt{\frac{c_{\textrm{IR}}J_\text{IR}}{6}}=\frac{\pi r_0}{2G_\text{N}}, \label{eq:CardyIR}
\end{equation}
and this is actually equal to the Bekenstein-Hawking entropy (\ref{eq:BH}).

Notice that $S_\text{IR} < S_\text{UV}$, which originates from the fact  $c_\text{IR} < c_\text{UV}$.
This is nothing but the celebrated Zamolodchikov's c-theorem in two dimensional field theory.
In the next section, we define the c-function from the gravity side, which explains the reason why
$S_\text{UV}$ gives the maximum possible entropy~\cite{car,nos,park}.


\section{Holographic Renormalization Group Flow}

\subsection{Review of Hamilton-Jacobi equation}

A key of the gauge/gravity correspondence is that the radial coordinate of the gravity theory is related to the
energy scale of the field theory on the boundary. Then renormalization group flow of the field theory is understood
from the gravity side as the variation of boundary values along the radial coordinate.
This is the so-called holographic renormalization group flow, and can be well analyzed by using Hamilton-Jacobi formalism.
Let us quickly review this formalism below.

Since the radial coordinate plays a special role, we reparametrize the metric so as to be an Euclidean 
ADM form\footnote{Here we employ the same notations for canonical variables, $N$, $N^i$ and so on, 
as in the previous section, rather than introducing new ones. Hopefully it might not make any confusion.}
\begin{equation}
ds^2=N^2d\rho^2+g_{ij}\left(dx^i+N^id\rho\right)\left(dx^j+N^jd\rho\right). \label{ADM}
\end{equation}
Here $\rho$ corresponds to the radial coordinate and $x^i$ parametrizes the two dimensional space-time.
In fact, as explained in the appendix \ref{sec:app}, the two dimensional metric and
the scalar field are written as
\begin{alignat}{3}
  g_{ij} = \frac{1}{\mu^2} \eta_{ij}, \qquad \phi = \frac{2 a r_0}{\rho}, \qquad
  \mu^2 \equiv \frac{a^4\ell^2}{\rho^2(e^{-a^2r_0^2/\rho^2}-e^{-a^2})}. \label{scale}
\end{alignat}
Note that $\mu \to 0$ when $\rho \to \infty$ and $\mu \to \infty$ when $\rho \to r_0$.
Since $\mu$ gives the length scale of the two dimensional field theory,
we see that UV region corresponds to the spatial infinity,
and IR region does to the horizon.

By inserting the ADM decomposition of the metric, the Lagrangian becomes
\ba
  \mc{L}_\text{E} &= \sqrt{-g} N \big( R^{(2)} - V(\phi) - K^{ij}K_{ij} + K^2 \big)
  - \frac{\sqrt{-g} }{2N} (\dot \phi - N^i \pa_i \phi)^2 
  - \frac{\sqrt{-g} N}{2} \pa_i \phi \pa^i \phi, \label{eq:Elag}
\end{alignat}
where the dot represents a derivative with respect to $\rho$, and 
$K_{ij} = \frac{1}{2N}(\dot g_{ij} - 2 \nabla_{(i} N_{j)})$.
$R^{(2)}$ stands for the scalar curvature made from the two dimensional metric $g_{ij}$.
Momenta conjugate to $g_{ij}$ and $\phi$ are given by $\pi^{ij} = - \sqrt{- g} (K^{ij} - g^{ij} K)$ and
$\pi_\phi = - \frac{1}{N} \sqrt{- g} (\dot \phi - N^i \pa_i \phi)$.
Up to total derivative terms, the Hamiltonian density is expressed as
$\mc{H}_\text{E} = \pi^{ij} \dot g_{ij} + \pi_\phi \dot \phi - \mc{L}_\text{E} = N \mathcal{H} + N^i \mathcal{P}_i$
in which $\mathcal{H}$ and $\mathcal{P}^i$ are defined by
\begin{align}
\frac{1}{\sqrt{-g}} \mathcal{H}&= \frac{1}{(-g)}\left(
(\pi^i_i)^2-(\pi_{ij})^2-\frac{1}{2}\pi_\phi^2 \right)
+V(\phi)-R^{(2)}+\frac{1}{2}(\partial_i\phi)^2,
\label{hamiltonian}\\
\frac{1}{\sqrt{-g}} \mathcal{P}^i&= -2\nabla_j\left(\frac{1}{\sqrt{-g}}\pi^{ij}\right)+\frac{1}{\sqrt{-g}}\pi_\phi\partial^i\phi.
\label{momentum}
\end{align}
It is apparent that $\mathcal{H}=\mathcal{P}^i=0$ since $N$ and $N^i$ are just the Lagrange multipliers.

Now let $\overline{g}_{ij}(x,\rho)$ and $\overline{\phi}(x,\rho)$ be the classical solutions of the bulk theory.
Then we denote the cut-off scale as $\rho_c$, and represent boundary values like
$\overline{g}_{ij}(x,\rho_c)=g_{ij}(x)$ and $\overline{\phi}(x,\rho_c)=\phi(x)$.
Substituting the classical solutions into the Lagrangian (\ref{eq:Elag}) and integrating over the three dimensions, 
we obtain a functional with respect to $g_{ij}$ and $\phi$.
We denote this functional as $S[g,\phi;\rho_c] = 
16 \pi G_\text{N} \mathcal{I}$.
Using the equations of motion, the variation of $S[g,\phi;\rho_c]$ with respect to 
$\rho_c$, $g_{ij}(x)$ and $\phi(x)$ is given by
\begin{equation}
  \delta S[g,\phi;\rho_c] = \frac{\partial S}{\partial \rho_c} \delta \rho_c
  + \int d^2x \frac{\delta S}{\delta \dot g_{ij}(x)} \delta g_{ij}(x) 
  + \int d^2x \frac{\delta S}{\delta \dot \phi(x)} \delta \phi(x).
\end{equation}
Combining this relation with 
$\frac{dS}{d\rho_c}=\int d^2x \mathcal{L}_\text{E}$, we find that the classical 
action is independent of $\rho_c$,
\begin{equation}
  \frac{\partial}{\partial \rho_c}S[g,\phi ; \rho_c]=
  - \int d^2x \,(N \mathcal{H} + N^i \mathcal{P}_i) = 0,
\end{equation}
and the boundary values of the conjugate variables are
\begin{align}
  \pi^{ij}(x) = \frac{\delta S}{\delta g_{ij}(x)}, \qquad
  \pi_\phi(x) = \frac{\delta S}{\delta \phi(x)}. \label{eq:mom}
\end{align}
Thus, the Hamilton-Jacobi equation reduces to only two constraints,
\begin{equation}
  \mathcal{H}\left(g_{ij}(x),\phi(x),\pi^{ij}(x),\pi_\phi(x)\right) = 0, \qquad
  \mathcal{P}^i\left(g_{ij}(x),\phi(x),\pi^{ij}(x),\pi_\phi(x)\right) =0,
\end{equation}
with eq.~(\ref{eq:mom}).
From the constraint $\mathcal{H} = 0$ one obtains the following equation,
\begin{equation}
  \frac{1}{(\sqrt{-g})^2}\left[-\left(g_{ij}\frac{\delta S}{\delta g_{ij}}\right)^2
  +\left(\frac{\delta S}{\delta g_{ij}}\right)^2
  +\frac{1}{2}\left(\frac{\delta S}{\delta \phi}\right)^2 \right] = V(\phi)-R^{(2)} + \frac{1}{2}(\partial_i\phi)^2.
\label{hjeq}
\end{equation}
As we will see later, it is possible to derive the conformal anomaly or the Callan-Symanzik equation from this equation.
The constraint $\mathcal{P}^i=0$ implies the invariance under the diffeomorphism of the theory in two dimensional 
space-time with $\rho$ fixed.


\subsection{Beta function and c-function from Hamilton-Jacobi equation}

Now let us solve the Hamilton-Jacobi equation (\ref{hjeq}).
First, since the bulk action diverges by taking $\rho_c\rightarrow \infty$, 
it is necessary to subtract such UV divergence.
For this purpose we divide the functional $S[g,\phi]$ into the local counter-term and the non-local part $\Gamma[g,\phi]$, 
which is the generating functional with respect to the external sources $g_{ij}(x)$ and $\phi(x)$.
Next we assign a weight $w$ to each variable such that $w=0$ for $g_{ij}(x),\,\phi(x)$ and $\Gamma[g,\phi]$ 
and $w=1$ for $\partial_i$.
From these assignment and the equation
$\delta\Gamma=\int d^2x(\delta g_{ij}(x)\delta\Gamma/\delta g_{ij}(x)+\delta \phi(x)\delta\Gamma/\delta \phi(x))$, 
$R^{(2)},\,\delta\Gamma/\delta g_{ij}(x)$ and $\delta\Gamma/\delta \phi(x)$ turn out to be $w=2$.

An integrand of the local counter-term with $w=0$ is written as a function of only the scalar field, $W(\phi)$,
and hence the classical action $S[g,\phi]$ is expressed 
as\footnote{It is possible to consider integrands of local 
counter-terms with $w=2$, such as $\Phi(\phi)R^{(2)}$ and 
$M(\phi)(\partial_i\phi)^2$, but these can be absorbed into 
the non-local term $\Gamma$ for the present case~\cite{fms1}.}
\begin{equation}
  S[g,\phi] = -\int d^2x\sqrt{-g}\,\big\{ W(\phi) + \cdots \big\} + 16\pi G_\text{N} \Gamma[g,\phi].
\end{equation}
The dots represent integrands of local counter-terms with $2 < w$.
Substituting this into (\ref{hjeq}) and comparing the terms with $w=0$, we obtain
\begin{equation}
  V(\phi)=-\frac{1}{2}W(\phi)^2+\frac{1}{2} \Big(\frac{\partial W(\phi)}{\partial \phi}\Big)^2. \label{weight0}
\end{equation}
The potential energy in the left hand side is given by eq.~(\ref{potential}), 
and the above equation is easily solved like
\begin{equation}
  W(\phi)=\frac{2}{a^2\ell}\left( \frac{\phi^2}{4}+1-e^{-a^2+\frac{\phi^2}{4}} \right). \label{super}
\end{equation}
In the range of $0\leq\phi\leq 2a$ (or $r_0 \leq \rho$), this function, often called ``superpotential",
monotonically increases as $\phi$ does. (See Fig.~2.)

\begin{figure}[tb]
\begin{center}
\begin{overpic}[width=8cm,clip]{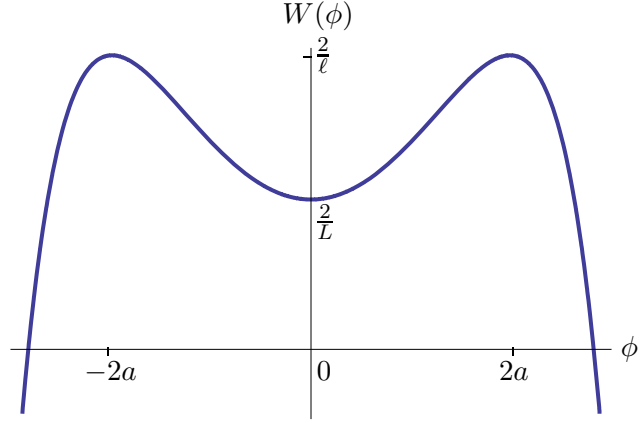}
  \put(231,24){$\phi$}
  \put(103,150){$W(\phi)$}
  \put(37,24){\line(0,1){3}}
  \put(190,24){\line(0,1){3}}
  \put(111,137){\line(1,0){3}}
  \put(28,15){$-2a$}
  \put(185,15){$2a$}
  \put(116,15){$0$}
  \put(115,72){$\frac{2}{L}$}
  \put(115,135){$\frac{2}{\ell}$}
\end{overpic}
\caption{Shape of the ``superpotential" $W(\phi)$.}
\end{center}
\end{figure}

From the terms with $w=2$ in eq.~(\ref{hjeq}), we obtain the following relation,
\begin{equation}
  \langle\, T^i_i(x) \,\rangle = -\frac{1}{8\pi G_\text{N}}\frac{1}{W(\phi)}R^{(2)}
  + \beta(\phi)\frac{1}{\sqrt{-g}}\frac{\delta \Gamma}{\delta \phi(x)}
  + \frac{1}{16\pi G_\text{N}}\frac{1}{W(\phi)}(\partial_i \phi)^2.
\label{trace}
\end{equation}
Here the energy momentum tensor is defined as
\begin{equation}
  \langle\, T^{ij}(x) \,\rangle=\frac{2}{\sqrt{-g}}\frac{\delta \Gamma[g,\phi]}{\delta g_{ij}(x)}, 
\end{equation}
and $\beta(\phi)$ is given by
\begin{equation}
  \beta(\phi)=\frac{2}{W(\phi)} \frac{\partial W(\phi)}{\partial \phi}. \label{beta}
\end{equation}
The notation $\beta(\phi)$ is adopted here, since it can actually be interpreted as the beta function for 
the dual field theory from the equation
\begin{alignat}{3}
  \mu \frac{d \phi}{d \mu} = 
  \frac{\phi(1-e^{-a^2+\frac{\phi^2}{4}})}{\frac{\phi^2}{4}+1-e^{-a^2+\frac{\phi^2}{4}}}
  = \beta(\phi).
\end{alignat}
The parameter $\mu$ in eq.~(\ref{scale}) is regarded as the scale of the two dimensional theory 
of $x^i$-space at a fixed $\rho$-slice, so it is really possible to identify $\beta(\phi)$ with the beta function.

Now we assume that the scalar field $\phi(x)$ is homogeneous on the two dimensional surface.
Then the third term in the right hand side of eq.~(\ref{trace}) becomes zero.
Furthermore, the second term vanishes at the points $\rho=\infty$ and $\rho=r_0$,
since $\beta(\phi) = 0$. Therefore we obtain,
\begin{equation}
  \langle\, T^i_i(x) \,\rangle \Big|_{\rho=\infty\textrm{ or } r_0}=
  - \frac{1}{24\pi} \frac{3}{G_\text{N} W(\phi)} R^{(2)} \Big|_{\rho=\infty\textrm{ or } r_0}. \label{anomaly}
\end{equation}
The vanishing of the beta function indicates that the two dimensional theory is conformally invariant, 
and the above equation corresponds to the conformal anomaly 
for the $\textrm{CFT}_2$ at UV ($\rho=\infty$) or IR ($\rho=r_0$).
We can read off these central charges as
$\frac{3}{G_\text{N}W(\phi)}|_{\rho=\infty} = c_\text{UV}$ and
$\frac{3}{G_\text{N}W(\phi)}|_{\rho=r_0}=c_\text{IR}$.
These agree with the results (\ref{cuv}) and (\ref{cir}) in the last section.
As a remark, it is found that (\ref{momentum}) leads to $\nabla_j \langle T^{ij} \rangle=0$.
This guarantees the absence of the gravitational anomaly.

From the calculation of the central charges for the $\textrm{CFT}_2$ (\ref{anomaly}), 
we can think of a function at any value of $\rho$,
\begin{equation}
  \mathcal{C}(\phi) = \frac{3}{G_\text{N}W(\phi)}. \label{cfunction}
\end{equation}
This is the so-called c-function for the dual field theory.
Because the function $W(\phi)$, which is given by (\ref{super}), is non-negative for $0 \leq \phi \leq 2a$, 
it is clear that
\begin{equation}
  \mu\frac{d\mathcal{C}(\phi)}{d\mu} = \beta(\phi) \frac{d \mathcal{C}}{d \phi}
  = - \frac{3 \beta(\phi)^2}{2 G_\text{N} W(\phi)} \leq 0.
\end{equation}
The equality is satisfied only at $\rho=\infty$ and $\rho=r_0$ where the dual theory becomes conformally invariant.
The monotonicity of this function (\ref{cfunction}) is consistent with the Zamolodchikov's c-theorem~\cite{z}.



It is more striking that the relation (\ref{trace}) obtained from the Hamilton-Jacobi equation 
implies the Callan-Symanzik equation for the two dimensional field theory.
Let us assume that $\Gamma[g,\phi]$ is the generating functional of the correlation function 
in which $\phi$ appears as an external field for a scaling operator $\mathcal{O}(x)$.
Then $n$ point function in the background of $g_{ij}$ and $\phi$ is given by
\begin{alignat}{3}
  \langle \mathcal{O}(x_1)\cdots \mathcal{O}(x_n) \rangle_{g,\phi}
  = \frac{1}{\sqrt{-g}} \frac{\delta}{\delta\phi(x_1)} \cdots \frac{1}{\sqrt{-g}} 
  \frac{\delta}{\delta\phi(x_n)}\Gamma[g,\phi],
\end{alignat}
and ordinary $n$ point function $\langle \mathcal{O}(x_1)\cdots \mathcal{O}(x_n) \rangle$ 
is obtained by setting $g_{ij}=\frac{1}{\mu^2}\eta_{ij}$ and $\phi=\phi(\rho)$ in the above equation.
By acting with $n$ functional derivatives 
$\frac{1}{\sqrt{-g}} \frac{\delta}{\delta\phi(x_1)} \cdots \frac{1}{\sqrt{-g}} \frac{\delta}{\delta\phi(x_n)}$
on eq.~(\ref{trace}), we obtain
\begin{alignat}{3}
  &\left[-2g_{ij}(x)\frac{\delta}{\delta g_{ij}(x)}+\beta(\phi(x))\frac{\delta}{\delta\phi(x)}
  \right] \langle \mathcal{O}(x_1)\mathcal{O}(x_2) \cdots \mathcal{O}(x_n) \rangle_{g,\phi} \notag
  \\
  &+\sum^n_{k=1}\delta(x-x_k) \frac{\partial \beta(\phi)}{\partial \phi}(x)
  \langle \mathcal{O}(x_1)\cdots\mathcal{O}(x_k)\cdots\mathcal{O}(x_n) \rangle_{g,\phi} = 
  \text{(two derivative terms)}. \label{cseq1}
\end{alignat}
Integrating this equation over two dimensional coordinate $x$ and setting $g_{ij}=\frac{1}{\mu^2}\eta_{ij}$ 
and $\phi=\phi(\rho)$, it becomes
\begin{equation}
  \left( \mu\frac{\partial}{\partial \mu} +\beta(\phi)\frac{\partial}{\partial \phi} -n\gamma(\phi) \right)
  \langle \mathcal{O}(x_1)\mathcal{O}(x_2)\cdots\mathcal{O}(x_n) \rangle = 0,
\label{cseq2}
\end{equation}
where $\gamma(\phi)=-\partial \beta(\phi)/\partial \phi$ is the anomalous dimension.
This is just the Callan-Symanzik equation of the two dimensional field theory.

In conclusion, with the help of the Hamilton-Jacobi formalism,
we have derived the central charges for $\textrm{CFT}_2$ at UV and IR regions from the conformal anomaly, 
and they are certainly connected by the c-function defined from the three dimensional gravity theory.
Our black hole solution interpolates two $\textrm{CFT}_2$ at the infinity and the horizon.



\section{Summary and Discussion}

In this paper, we constructed the new extremal black hole solution in the three dimensional gravity theory 
coupled to a single scalar field, and investigated AdS$_3$/CFT$_2$ correspondences which are 
realized at the spatial infinity and the horizon.

The black hole solution (\ref{eg}) which we have found is obtained by 
choosing the potential as in eq.~(\ref{potential}).
The potential takes extrema at $\phi=0$ and $\phi=2a$, which correspond to 
the spatial infinity and the horizon, respectively.
Around the spatial infinity and the horizon, the metric approaches locally AdS$_3$ geometries with the radius $L$ and $\ell$.
This means that the near horizon region corresponds to the IR fixed point of two dimensional field theory, 
and the spatial infinity does to the UV one.
Thus the solution represents the $\textrm{CFT}_2$-interpolating black hole.

In order to confirm the above feature, we showed that the Virasoro algebras surely exist at the spatial infinity and the horizon
by employing the canonical formulation of the gravity theory.
Near the infinity we used the usual boundary condition (\ref{hypersurface bd}). 
On the other hand, near the horizon we took near horizon limit (\ref{nearh}) and put the boundary condition (\ref{bchorizon}).
The central charges are given by $c_\text{UV}$ (\ref{cuv}) and $c_\text{IR}$ (\ref{cir}), 
and they are related to the corresponding depths of the potential (\ref{potential}).
It is easy to see the inequality $c_\text{IR} < c_\text{UV}$, which is consistent with Zamolodchikov's c-theorem.

From the viewpoint of $\textrm{CFT}_2$ at UV region, the spatial infinity of the black hole is represented as
excitations of only left moving modes. The energy of the geometry is given by the sum of the black hole mass and 
the energy of scalar field, which will be interpreted as that of a domain wall.
By using the Cardy's formula, the entropy was estimated as in eq.~(\ref{eq:CardyUV}).
From the viewpoint of $\textrm{CFT}_2$ at IR region, the near horizon geometry of the black hole also corresponds to
excitations of only left moving modes. The entropy (\ref{eq:CardyIR}) calculated by using Cardy's formula precisely 
agrees with the Bekenstein-Hawking entropy of the black hole (\ref{eq:BH}).
As a result, we have obtained the relation $S_\text{BH} = S_\text{IR} < S_\text{UV}$,
which originates from the c-theorem and clearly explains the reason why maximum possible entropy 
conjecture holds.

Furthermore we have investigated the renormalization group flow of two dimensional field theory from the gravity side.
This is the so-called holographic renormalization group flow, and the Hamilton-Jacobi formalism played an important role
for the analyses.
In fact, by using this formalism, we have derived the flow equation (\ref{hjeq}) along each radial surface.
The flow equation is solved order by order with respect to the weight, and we obtained eq.~(\ref{trace}) or eq.~(\ref{anomaly})
which expresses the conformal anomaly for the $\textrm{CFT}_2$ at the critical points $\rho=\infty$ or $\rho=r_0$.
Eq.~(\ref{beta}) is identified with the beta function of dual field theory, and the c-function is defined 
as eq.~(\ref{cfunction}). The c-function is monotonically decreasing along the flow from UV to IR,
and Zamolodchikov's c-theorem is satisfied.
Finally, the Callan-Symanzik equation (\ref{cseq2}) was derived for the two dimensional field theory 
dual to the bulk theory of gravity.
The conclusion is that two $\textrm{CFT}_2$ satisfying the Virasoro algebra do live on two 
boundaries of our black hole solution, and these are connected via the holographic renormalization group flow.

Recently there are many discussions on the consistency of the $\textrm{CFT}_2$ on the boundary of 
$\textrm{AdS}_3$~\cite{witten2}-\cite{s}.
The holographic perspective, such as our solution which connects different $\textrm{CFT}_2$, will become more useful.
Especially, $c_{\textrm{IR}} < c_{\textrm{UV}}$ means that massless modes in the UV region becomes massive 
and are integrated out while the energy scale is decreasing.
Although our system is one of toy models, we will have to consider how this process is explained in the consistent 
two dimensional field theory.

Moreover, it is an interesting problem to investigate how $\textrm{CFT}$-interpolating black holes, such as ours, can be embedded into higher dimensional gravity or superstring theory. 
It is known that the BTZ black hole is embedded into black ring
 solutions or the so-called M5 system~\cite{msw}.
For example in five dimensions, we can realize the asymptotically flat multi-centered black ring solutions, for which each near horizon  geometry is $\textrm{AdS}_3$.
It is suggested that  such multi-centered solutions represent the decay of branes in the holographic 
viewpoint~\cite{dbdesmb}.
In three dimensions there is also the multi-centered BTZ~\cite{ms}.
The problems about asymptotically AdS multi-centered black holes are 
left open for the future work.


\section*{Acknowledgements}

YH would like to thank Satoshi Iso for useful discussions.
The authors would like to thank Akihiro Ishibashi for helpful communications.
We also thank the Yukawa Institute for Theoretical Physics at Kyoto University, 
where this work was developed during the YITP-W-08-04 on 
``Development of Quantum Field Theory and String Theory''.
KH is supported in part by JSPS Research Fellowship for Young Scientists. 
The work of YH is partially supported by the Ministry of Education, Science, 
Sports and Culture, Grant-in-Aid for Young Scientists (B), 19740141, 2007.

\appendix

\section{Supplementary Calculations on Central Extensions}

\subsection{At the spatial infinity}\label{appb1}

Here we give supplementary calculations at the spatial infinity on the central extension for left moving modes.
In this subsection, we denote the metric (\ref{asymptotic}) as $\bar{G}_{\mu\nu}$, and
other quantities with ``bar" means that they consist of $\bar{G}_{\mu\nu}$.

In order to evaluate the central extension, we need to calculate the explicit form of eq.~(\ref{hypersurface bd}).
By using the Killing vector (\ref{eq:Killing}), it is given by 
$G_{\mu\nu} = \bar{G}_{\mu\nu} + \bar{D}_\mu \bar{\xi}^+_{n\,\nu} + \bar{D}_\nu \bar{\xi}^+_{n\,\mu}$, 
and the result becomes
\begin{alignat}{3}
  G_{\mu\nu} &= 
  \begin{pmatrix}
    - \frac{r^2}{L^2} + \frac{2 r_0^2}{L \ell} & 0 & \frac{r_0^2}{\ell}
  \\
    0 & \frac{L^2}{r^2} + \frac{2 e^{-a^2} r_0^2 L^3}{r^4 \ell} & 0
  \\
    \frac{r_0^2}{\ell} & 0 & r^2 
  \end{pmatrix} \notag
  \\
  &\quad\, + e^{i n x^+} 
  \begin{pmatrix}
    \frac{1}{2} i n^3 + \frac{3 i r_0^2 n}{L \ell} &
    \frac{(1+2 e^{-a^2}) n^2 r_0^2 L^2}{2 r^3 \ell } & 
    \frac{1}{2} i L n^3 + \frac{2 i r_0^2 n}{\ell} 
  \\[0.1cm]
    \frac{(1+2e^{-a^2}) n^2r_0^2 L^2}{2r^3 \ell} & 
    \frac{2 i e^{- a^2} n r_0^2L^3}{r^4 \ell } & 
    \frac{(1+2e^{-a^2}) n^2 r_0^2 L^3}{2r^3 \ell } 
  \\[0.1cm]
    \frac{1}{2} i L n^3 + \frac{2 i r_0^2 n}{\ell} & 
    \frac{(1+2e^{-a^2}) n^2 r_0^2 L^3}{2r^3 \ell } &
    \frac{1}{2} i L^2 n^3 + \frac{i L r_0^2 n}{\ell}
  \end{pmatrix}.
\end{alignat}
Note that the variation contains the factor $e^{inx^+}$ because $\xi^+_{n\,\mu}$ just represents one Fourier mode of
the general coordinate transformation.
From the ADM decomposition (\ref{eq:ADM}), the two dimensional variables are evaluated as
\begin{alignat}{3}
  g_{ij} &= 
  \begin{pmatrix}
    \frac{L^2}{r^2} + \frac{2 e^{-a^2} r_0^2 L^3}{r^4 \ell} & 0
  \\
    0 & r^2 
  \end{pmatrix} 
  + e^{i n x^+} 
  \begin{pmatrix}
    \frac{2 i e^{- a^2} n r_0^2L^3}{r^4 \ell } & 
    \frac{(1+2e^{-a^2}) n^2 r_0^2 L^3}{2r^3 \ell } 
  \\
    \frac{(1+2e^{-a^2}) n^2 r_0^2 L^3}{2r^3 \ell } &
    \frac{1}{2} i L^2 n^3 + \frac{i L r_0^2 n}{\ell}
  \end{pmatrix}, \notag
  \\
  N^r &= e^{i n x^+} \frac{(1 + 2e^{-a^2}) n^2 r_0^2}{2 \ell r}, \qquad
  N^\varphi = \frac{r_0^2}{\ell r^2} + i e^{i n x^+} \frac{\ell L n^3 + 4 r_0^2 n}{2 \ell r^2}, 
  \\
  N &= \frac{r}{L} - \frac{r_0^2}{\ell r} 
  - i n e^{i n x^+} \frac{6 r_0^2+L n^2\ell}{4 \ell r}, \qquad
  \phi = \frac{2a r_0}{r} + in e^{inx^+} \frac{a r_0}{r} \notag
\end{alignat}
The conjugate momenta are calculated as
\begin{alignat}{3}
  \pi^{ij} &= 
  \begin{pmatrix}
    0 & \frac{r_0^2}{\ell r^2}
    \\
    \frac{r_0^2}{\ell r^2} & 0
  \end{pmatrix}
  + e^{i n x^+} 
  \begin{pmatrix}
    - n^2 \frac{ L n^2 \ell + (3 - 2 e^{-a^2}) r_0^2}{2 \ell r} & 
    i \frac{ L \ell n^3 + 4n r_0^2}{2 \ell r^2}
    \\
    i \frac{ L \ell n^3 + 4n r_0^2}{2 \ell r^2} & 
    - \frac{(1 + 2e^{-a^2}) L^2 n^2 r_0^2}{\ell r^5} 
  \end{pmatrix},
  \\
  \pi_\phi &= e^{i n x^+} \frac{2 a (1 + e^{-a^2}) L n^2 r_0^3}{\ell r^4}. \notag
\end{alignat}
The boundary behaviors in eq.~(\ref{eq:infbc}) are consistent with 
these equations.

\subsection{At the horizon}\label{appb2}

Here we give supplementary calculations at the horizon on the central extension for left moving modes.
In this subsection, we denote the metric (\ref{tilde}) as $\bar{G}_{\mu\nu}$, and
other quantities with ``bar" means that they consist of $\bar{G}_{\mu\nu}$.

In order to evaluate the central extension, we need to calculate the explicit form of eq.~(\ref{bchorizon}).
By using the Killing vector (\ref{eq:Killing2}), it is given by 
$G_{\mu\nu} = \bar{G}_{\mu\nu} + \bar{D}_\mu \bar{\xi}^+_{n\,\nu} + \bar{D}_\nu \bar{\xi}^+_{n\,\mu}$, 
and the result becomes
\begin{alignat}{3}
  G_{\mu\nu} &= 
  \begin{pmatrix}
    - \frac{\ell^2}{4} & 0 & -\frac{r_0 \ell}{2} \tilde{r} 
    \\
    0 & \frac{\ell ^2}{4(1+\tilde{r}^2)} & 0
    \\
    - \frac{r_0 \ell}{2} \tilde{r} & 0 & r_0^2
  \end{pmatrix}
  + e^{i n \tilde{\varphi}}
  \begin{pmatrix}
    0 & 0 & 0
    \\
    0 & - \frac{i n \ell^2}{2 (1+\tilde{r}^2)^2} & \frac{n^2 \tilde{r} \ell^2}{4 (1+\tilde{r}^2)} 
    \\
    0 & \frac{n^2 \tilde{r} \ell ^2}{4 (1+\tilde{r}^2)} & 2 i n r_0^2
\end{pmatrix}.
\end{alignat}
Note that $\delta G_{\tilde{\varphi}\tilde{\varphi}}$ is just a leading contribution.
In such a case, the nonlinear terms become important and we need to employ the covariant formulation, 
in which the variation of the charge is defined as
\begin{alignat}{3}
  \delta_\eta Q^{C}[\xi] &= \int d\tilde{\varphi} \, \epsilon_{\mu\nu\tilde{\varphi}}
  \Big\{ \bar{\xi}^\nu \bar{D}^\mu \delta_\eta G^\sigma{}_\sigma 
  - \bar{\xi}^\nu \bar{D}_\sigma \delta_\eta G^{\mu\sigma} 
  + \bar{\xi}_\sigma \bar{D}^\nu \delta_\eta G^{\mu\sigma} 
  + \frac{1}{2} \delta_\eta G^\sigma{}_\sigma \bar{D}^\nu \bar{\xi}^\mu \notag
  \\
  &\qquad\qquad\qquad
  - \delta_\eta G^{\nu\sigma} \bar{D}_\sigma \bar{\xi}^\mu + \frac{1}{2} \delta_\eta G^{\nu\sigma} 
  (\bar{D}^\mu \bar{\xi}_\sigma + \bar{D}_\sigma \bar{\xi}^\mu) \Big\}. \label{eq:cov}
\end{alignat}

\section{Coordinate Transformation}\label{sec:app}

We give coordinate transformations which make the solution (\ref{metric}) with (\ref{eg}) 
into the Euclidean ADM form (\ref{ADM}).
Actually this can be done as follows,
\begin{alignat}{3}
  ds^2 &= e^{2h(r)}dr^2 + \frac{r^2}{a^4\ell^2} \left[d\theta^2-2(e^{-a^2r_0^2/r^2}-e^{-a^2})dtd\theta\right] \notag
  \\
  &= N^2d\rho^2+\frac{1}{\mu^2}\left[-(d\tau+N^\tau d\rho)^2+(d\sigma+N^\sigma d\rho)^2 \right]. \label{metric2}
\end{alignat}
In the first line, we defined $\theta\equiv a^2\ell\varphi+(1-e^{-a^2})t$. 
And in the second line we made the coordinate transformation of
\begin{align}
\rho&=r,\notag\\ \tau&=\frac{1}{4}\left( \frac{\theta}{e^{-a^2r_0^2/r^2}-e^{-a^2}}-2t
\right)-\theta,\notag\\ \sigma&=\frac{1}{4}\left( \frac{\theta}{e^{-a^2r_0^2/r^2}-e^{-a^2}}-2t \right)+\theta. \label{tausigma}
\end{align}
The two dimensional coordinate is denoted by $x^i=\tau,\sigma$. 
The functions $\mu$, $N$ and $N^i$ are written in terms of $(\rho,\tau,\sigma)$ as
\begin{align}
\mu^2&=\frac{a^4\ell^2}{\rho^2(e^{-a^2r_0^2/\rho^2}-e^{-a^2})}, \notag \\
N^2&=e^{2h(\rho)}=\frac{a^4\ell^2}{\rho^2}\left[1-e^{a^2 \left(r_0^2/\rho^2-1\right) } \right]^{-2},  \\
N^\tau&=N^\sigma=\frac{a^2r_0^2e^{-a^2r_0^2/\rho^2}}{4\rho^3(e^{-a^2r_0^2/\rho^2}-e^{-a^2})^2}(\sigma-\tau). \notag
\end{align}
In (\ref{metric2}) the coefficient $\mu$ can be regarded as the scale of the two dimensional theory of $(\tau, \sigma)$-space 
at a certain $\rho$-slice. 
Since  $\rho\rightarrow r_0$ means $\mu\rightarrow\infty$, let us call it the IR-region.
On the other hand, $\rho\rightarrow\infty$ or $\mu\rightarrow0$ indicates the UV-region.
Also notice that the scalar field becomes $\phi=\phi(\rho)$.

If $t$ and $\varphi$ are tuned, we are able to choose arbitrary $(\tau, \sigma)$ at any radius $\rho$ except for $\rho=r_0$.
But be careful of the transformation (\ref{tausigma}) at the horizon. 
For example, from the near horizon limit (\ref{nearh}),
\begin{equation}
\tau=\frac{e^{a^2}}{4}\left(\frac{\ell r_0}{2}\frac{\varphi'}{r'}-2t'\right)\frac{1}{\epsilon}
-a^2\ell\varphi'
\label{nearhtau}
\end{equation}
is naively seen to be divergent as $\sim\mathcal{O}(1/\epsilon)$.
However, since $t', r'$ and $\varphi'$ can be chosen arbitrary 
after rewriting (\ref{nearh}), $\tau$ can be kept finite and 
taken arbitrary if the inside of the brackets of (\ref{nearhtau}) 
is fine-tuned as $\sim\mathcal{O}(\epsilon)$ by the appropriate 
$t', r'$ and $\varphi'$.


\end{document}